\documentclass{aa}  
\usepackage{graphicx}
\usepackage{txfonts}
\usepackage{lipsum}
\usepackage{subcaption}         
\usepackage{booktabs}
\usepackage{lscape}             
\usepackage{placeins}           
                                
\usepackage[hypertexnames=false,colorlinks=true,linkcolor=blue,citecolor=blue]{hyperref}
\newcommand{\n}[1]{\textbf{[NOTE: #1]}}

\newcommand{\gaia}{\textit{Gaia}~}
\newcommand{\tgaia}{\textit{Gaia}}
\newcommand{\fmast}{f_{m,\text{ast}}}
\newcommand{\mjup}{\,M_\text{Jup}}
\newcommand{\msun}{\,M_\odot}

\newcommand{\di}{\text{d}}

\newcommand{\prob}[2]{\text{Pr}(#1 \mid #2)}
\newcommand{\likelihood}[2]{\mathcal{L}(#1 \mid #2)}

\begin{document}

   \title{Brown dwarfs and stars are dynamically indistinguishable as companions in AU-separation binaries}

%
%
%

 \author{Ben Pennell\inst{1,2}\thanks{Corresponding author: bepennell@mpia.de} \and
          Johanna M\"uller-Horn\inst{1,2} \and
          Sahar Shahaf\inst{1} \and
          Hans-Walter Rix\inst{1} \and
          Kareem El-Badry\inst{3,1} \and
          Jiadong Li\inst{1}}

   \institute{
    Max-Planck-Institut für Astronomie, Königstuhl 17, 69117 Heidelberg, Germany
    \and Fakultät für Physik und Astronomie, Universität Heidelberg, Im Neuenheimer Feld 226, 69120 Heidelberg, Germany \and
    Department of Astronomy, California Institute of Technology, Pasadena, CA 91125, USA
}
\date{Submitted 16 November 2026}

\abstract{Brown dwarfs occupy the mass range between stars and planets (${\sim}\!13$--$80\,M_\mathrm{Jup}$), populations whose dominant formation channels differ, which leave measurable differences in the orbital eccentricity and host-star metallicity distributions. We assemble a sample of astrometric binaries with main sequence primaries and brown dwarf companions at $\mathcal{O}(\mathrm{AU})$ separations from {\it Gaia} DR3. The observable photocenter motions for such binaries are degenerate between a genuinely low-mass companion and a near-equal-mass binary. We break this degeneracy using {\it Gaia} XP spectrophotometry and follow-up FEROS spectroscopy, which removes removes ${\sim}70\%$ of initial low-mass-function candidates as likely near-twin binaries and leaves residual contamination of only a few percent. The resulting 368 systems, with masses and metallicities determined from the XP spectra and a selection function forward-modelled with {\tt Gaiamock}, form the largest homogeneous sample of brown dwarf companions with astrometric orbits. We find the eccentricity distribution to be well described by a normal distribution with $\bar{e}=0.46\pm0.02$ and $\sigma_e=0.16\pm0.01$. This is comparable to eccentricities of stellar companions at the same periods and higher than those of giant planets, with no significant dependence on companion mass ($\alpha=-0.13\pm0.09$) or period ($\beta=0.10\pm0.05$). The host-star metallicity distribution is likewise stellar-like, extending to $\mathrm{[M/H]}\lesssim-0.5$, in contrast to the predominantly super-solar metallicity hosts of giant planets. After accounting for selection, the companion-mass distribution over all brown dwarf masses is consistent with an extension of stellar-binary demographics. Brown dwarf companions at AU separations down to at least ${\sim}20\,M_\mathrm{Jup}$ therefore appear to form predominantly through a binary-star-like channel. A possible transition to planet-like formation at lower masses cannot be located with the present sample.}


\maketitle
\nolinenumbers

\section{Introduction}
Binary star formation appears to have two dominating modes: direct fragmentation of a collapsing cloud with $\mathcal{O}(1000\mathrm{AU})$ separations \citep{Adams1989, Kratter2016} and circumstellar disk instability with $\mathcal{O}(100\mathrm{AU})$ separations \citep{Fisher2004, Kratter2016}. The latter process appears to dominate as stellar multiplicity is most common at this scale \citep[corresponding to ${\sim}10^5$\,days,][]{MoeDiStefano2017}. Meanwhile, planets appear to form through core-accretion in a protoplanetary disk \citep{Pollack1996}, or perhaps through the binary-star like disk instability \citep{Kratter2016}.
To create close binaries, the companion must then migrate inwards through interactions with the remaining debris, companions, or passing systems \citep{Tohline2002, Moe2018, Offner2023}.
We generally observe binaries after dynamical processing has ceased, where systems have become decoupled from their birth environments and their orbit distributions reflect both the formation process and dynamical processing.

This makes the overlapping mass regime between planets and stars -- brown dwarfs -- both challenging to probe and rich in physics.
Brown dwarfs are conventionally defined as too small to be called stars, not able to sustain core-hydrogen burning \citep[${\lesssim}80\mjup$,][]{Hayashi1963, Kumar1963, Kirkpatrick2005}, but massive enough for deuterium burning, which does not happen below ${\sim}13\mjup$ \citep{Spiegel2011, Burrows1997, Boss2007}.
Somewhere in or below the brown dwarf mass range is where formation changes between forming ``bottom-up''  from core accretion within protoplanetary discs (forming ``like a planet'') or through the fragmentation and direct collapse of a star forming cloud or disk (forming ``like a star''), and it has been proposed to separate brown dwarfs from planets based on formation mechanism rather than mass \citep[e.g.][]{Black1997}.
Previously, dearths of intermediate mass brown dwarfs in binaries (${\sim}30-50\mjup$) have been reported as a ``brown dwarf desert`` \citep{GretnerLineweaver06, MaGe14, Stevenson+23} representing such a turnover in formation mechanism, but this remains unclear as others have found no such dearth \citep{GaiaCollaboration2023, Wallace2026}.
It is worth creating a large, well understood sample of brown dwarfs in binaries to address where this turnover may exist.

Different formation histories may be hard to discern in individual objects, but should be reflected in population properties.
For example, at AU-separations, the distribution of orbital eccentricities is typically smaller in giant planets $\bar{e}{\sim}0.2$ \citep{Bowler2020, Bitsch2020} than stellar companions $\bar{e}{\sim}0.4$ \citep{Duquennoy1991, Wu+2025}.
Additionally, giant planets appear to form predominantly in systems of super-solar metallicity \citep{Gan2025}, while the distribution of metallicities in stellar binaries is anti-correlated with metallicity at these separations \citep{Raghavan+2010, Moe2019, El-Badry2019}.
Astrometric binaries with main sequence stars and brown dwarf companions allow for the identification of the orbital parameters of the orbit (period, eccentricity), while spectroscopic measurements of the system determines the properties of the primary (mass, metallicity). Since binaries at ${<}\mathcal{O}(10\text{AU})$ separations likely formed from the same cloud core, the metallicity of a brown dwarf companion is well approximated by that of the primary.

The \textit{Gaia} mission \citep{Gaia} and its third data release \citep[DR3,][]{DR3} provides an unprecedented opportunity to construct such a sample with its 34 month baseline and multiple yearly precise astrometric, photometric, and spectroscopic measurements.
168~000 objects with AU-scale orbits passing stringent cuts have a full orbit solution published in \tgaia's non single star (NSS) catalogue \citep{Halbwachs+2023}.
For main sequence stars, their mass and metallicity can be readily obtained through {\it Gaia}'s low-resolution BP-RP ``XP'' spectra \citep{Li+2025}, which is available homogeneously for every object brighter than $G{<}17.65$ in \textit{Gaia}'s G-band \citep{DeAngeli+2023}. 
Determining the mass of the brown dwarf requires handling the degeneracy where low-mass dark companions and near-twins may lead to identical photocentre motions, which can be resolved spectroscopically \citep{Li+2025}.
Our sample constructed this way contains 368 main sequence-brown dwarf (MS-BD) systems, the largest sample of astrometric binaries with brown dwarf companions.
Finally, such a catalogue has a known selection function which can be forward modelled \citep{ElBadry+2024}, providing unprecedented access to the underlying physics.

The remainder of this paper is organised as follows: Section~\ref{sec:Sample} describes the creation of a sample both pure and complete in brown dwarf companions with full orbit solutions, and summarises follow-up observations which confirm the contamination rate of impostors is low.
Section~\ref{sec:Gaiamock} describes how \texttt{Gaiamock} can be used to untangle selection effects on the sample.
The distributions of eccentricities and metallicities are compared to planetary and stellar companions in Sect.~\ref{sec:Eccentricities} and \ref{sec:Metallicities} respectively.
We discuss our results in Sect.~\ref{sec:discussion}, and then summarise and conclude them in Sect~\ref{sec:Conclusion}.

\section{A sample of brown dwarfs} \label{sec:Sample}
\subsection{Sample creation} 
\begin{figure}
    \centering
    \includegraphics[width=\linewidth]{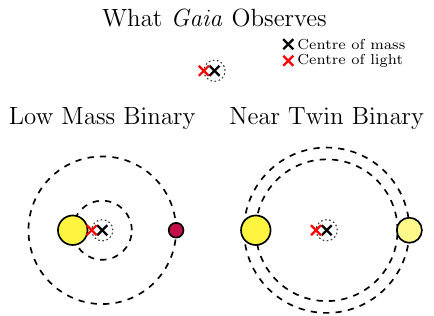}
    \caption{A demonstration of how astrometric wobbles are inherently degenerate as near-twin and low-mass binaries can each generate the same astrometric signal.}
    \label{fig:DegeneracyDemonstration}
\end{figure}
For a main sequence primary, a brown dwarf companion would induce a small on-sky astrometric wobble that may be detected by \textit{Gaia}. 
For a series of astrometric measurements of a spatially unresolved binary, the astrometric signal comes from tracking the motion of the centre of light as it orbits the centre of mass.
It is convention to describe the strength of an astrometric signal with the mass function $\fmast$, which is related to the mass of the companion based on observables: the orbital period $P$ and photocentre semi-major axis $a_0$ scaled by the parallax $\varpi$, via Kepler's third law:
\begin{equation}
    \fmast = \biggl(\frac{a_0}{\varpi}\biggr)^3\biggl(\frac{P}{1\,\text{yr}}\biggr)^{-2}.
\end{equation}

Where the photocentre semi-major axis is the semi-major axis of the ellipse of the orbit of the centre of light about the centre of mass, which is related to the true semi-major axis $a$ for a binary with a mass ratio $q=M_2/M_1$ and flux ratio $f=f_2/f_1$ \citep{Penoyre2020}
\begin{equation}
    a_0 = \biggl(\frac{q}{1+q} - \frac{f}{1+f}\biggr)~a.
\end{equation}
Which shows that astrometric measurements are inherently degenerate, as both a veritably small companion (low $q$, low $f$) and a near twin ($q\rightarrow1, f\rightarrow1$) companion can in principle generate the same astrometric signal, which is shown schematically in Fig.~\ref{fig:DegeneracyDemonstration}.

Fortunately, this degeneracy can be resolved through spectroscopy.
\textit{Gaia} DR3 published low ($R{\sim}$20--100) resolution spectra from its 330--1050\,nm photometers for 220 million sources \citep{DR3, DeAngeli+2023}.
\cite{Li+2025} used the XP spectral energy distribution (SED) to identify single stars by fitting single-star and binary spectral models and comparing the goodness-of-fit.
For a brown dwarf companion to a main sequence star, the observed SED should be nearly unchanged due to the negligible flux contribution of the brown dwarf.
Conversely, a near-twin binary should reveal itself in the observed SED as it will manifest as being over-luminous for the apparent SED, as both stars contribute light of a similar temperature.

Of the $168~000$ orbit solutions published in \tgaia DR3 \citep{Halbwachs+2023}, we assembled all those with \gaia XP spectra published between $50$ and $300$\,pc with an absolute magnitude dimmer than $G<4.5$, to reduce our sample to M-, K-, and G- type primaries. This resulted in $24~815$ sources, of which $9309$ have a mass function below $10^{-2}\msun$.
We then retained sources that \cite{Li+2025} determine are most likely to be apparently single, or consistent with a brown dwarf-mass companion following \cite{Pennell+2026}, leaving $11~124$ sources, of which $3956$ have a mass function below $\fmast\leq10^{-2}\msun$.

A result of constructing the sample in this manner is that we have excluded systems with companions which contribute significantly to the flux, which means we can determine precise masses and metallicities of the primary star based on its published XP-spectra, which has uncertainties around $10\%$ \citep{Li+2025} near the main sequence.
Astrometry provides the opportunity to obtain precise companion mass estimates, and is not limited to lower bounds as in radial velocity surveys.
To determine the companion mass and uncertainty, we sampled orbital parameters from the published best-fit parameters and correlation matrix $1000$ times, and under the assumption the companion is invisible, we computed the resulting companion mass.
We recorded the companion masses as the mean computed mass, with the uncertainty representing the central 68\% ($1$-$\sigma$) of the computed masses, which results in a typical companion mass uncertainty below $10\%$.
To restrict ourselves to the brown dwarf companion regime, we retain only sources with companion masses assigned to be under the conventional $80\mjup$ limit, which results in 379 MS-BD systems.

\subsection{Sample purity} \label{sec:Contamination}
\begin{figure}
    \centering
    \includegraphics[width=\linewidth]{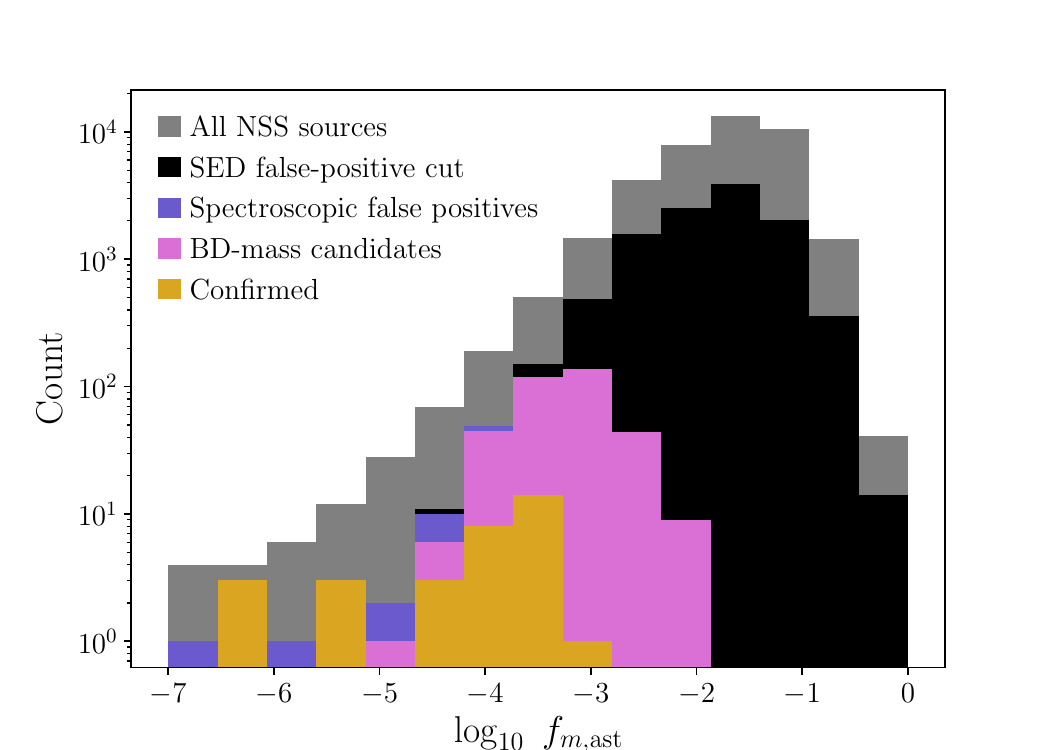}
    \caption{stacked distribution of the astrometric mass function $\fmast$ for the sample, showing all sources with astrometric solutions in \gaia which are assigned an XP spectra within 300\,pc shown in grey. Then, the subset of sources which pass the spectroscopic cut from \cite{Li+2025} aimed to eliminate near-twin false positives are shown in black. Finally, we show the sources which enter our sample (pink) based on the estimated companion mass, and show the confirmed companions (gold) and known erroneous orbits (blue).}
    \label{fig:massfunction}
\end{figure}

\begin{figure*}
    \centering
    \includegraphics[width=\linewidth]{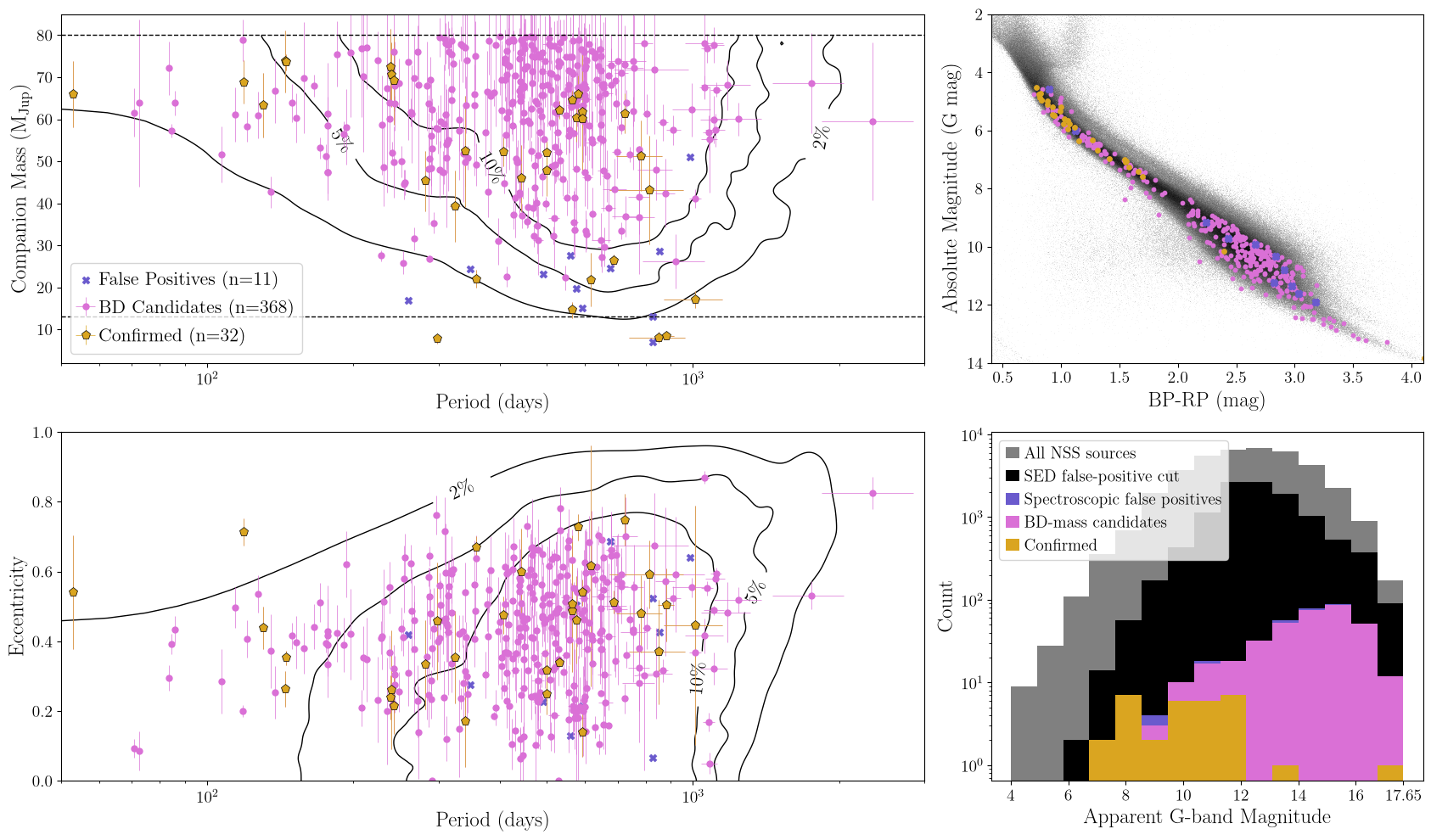}
    \caption{Visual description of the MS-BD sample. Left panels: brown dwarf masses (top) and eccentricities (bottom) as a function of orbital period for our entire sample (pink, Sect.~\ref{sec:Sample}), with confirmed brown dwarfs and false positives found in our follow-up and previous surveys highlighted (Sect.~\ref{sec:Contamination}). Black contours correspond to detection probabilities that are marginalised over distance and primary mass as computed in Sect.~\ref{app:desert}, and it can be seen that the sample roughly matches the expected distribution from selection effects. Top right panel: MS-BD sample overlayed on the \gaia colour-magnitude diagram, with the density of all sources within 300\,pc with XP spectra shown in greyscale. Bottom right panel: stacked distribution of the apparent \gaia G-band magnitude for the sample, showing each step in the selection similar to Fig.~\ref{fig:massfunction}.}
    \label{fig:SampleDiagnostics}
\end{figure*}

While the \textit{Gaia} XP-based spectroscopic vetting can reliably identify and reject near-twin binaries through their composite spectral signatures, some contamination may remain in the cleaned BD sample. We assess the level of residual contamination (i) through spectroscopic follow-up of BD candidates, testing the consistency between their astrometric orbits and measured radial velocities, and (ii) by compiling results from previous follow-up studies.

Radial velocities provide an independent way to break the flux-ratio degeneracy inherent to astrometric measurements. For near-twin binaries, the expected radial-velocity semi-amplitude for AU-scale orbits is $\mathcal{O}(10~\text{km/s})$, whereas those of main-sequence stars with brown-dwarf companions at these separations are typically of order $\mathcal{O}(1~\text{km/s})$. Therefore, given a \textit{Gaia} astrometric orbit solution and an estimate of the primary mass, we can predict the radial velocity curves under the two scenarios. Even a small number of radial velocities can then distinguish between low-mass and near-twin companions.

We obtained follow-up spectra with the Fiber-fed Extended Range Optical Spectrograph \citep[FEROS,][]{Kaufer1999} mounted on the MPG/ESO 2.2m telescope at La Silla Observatory (Program P117.2AK3, PI Müller-Horn). FEROS has a resolving power $R \approx 50000$ and covers a wavelength range approximately 3500–9200\,\AA. We observed 30 of our sources with the lowest mass functions that were accessible from La Silla, obtaining between one and three epochs per source (see Table~\ref{tab:followedupbds}).
The spectra were reduced with the CERES pipeline \citep{Brahm2017}, which also provides radial velocity measurements based on template cross-correlation.

The astrometric orbital solution, together with the adopted primary mass, determines the expected radial-velocity variation but leaves the systemic velocity unconstrained. We therefore use the \textit{Gaia} DR3 mean radial velocity as an estimate of the systemic velocity when comparing the predicted and observed radial velocities. Its typical uncertainty is of order a few $\,\mathrm{km\,s^{-1}}$. For near-twin binaries, the predicted radial-velocity amplitudes are much larger than this uncertainty, such that even a single measurement can reveal a significant deviation from the expected systemic velocity, provided it is obtained at a suitable orbital phase. For brown dwarf companions, in contrast, the predicted amplitudes are often comparable to the uncertainty in the systemic velocity, making their confirmation considerably more challenging with sparse temporal sampling.

We identify one clear false positive among the 30 followed-up sources. \textit{Gaia} DR3 5213765361595890304, whose radial velocities are inconsistent with the brown dwarf companion scenario and instead indicate a near-twin binary (Fig.~\ref{fig:follow-up}, Appendix). For the remaining 29 sources, we find no clear evidence of the large radial-velocity variations expected for near-twin companions. However, given the limited number of epochs (one to three per source), the available measurements are generally insufficient to validate the \textit{Gaia} astrometric orbit solutions or to rule out spurious solutions.
For a few sources, the measured radial velocities are not only inconsistent with the near-twin scenario but are also better matched by the predicted orbit of a brown dwarf companion than by a constant radial velocity. These cases provide tentative support for the presence of low-mass companions, although the sparse temporal sampling prevents a robust confirmation. One such example is shown in Fig.~\ref{fig:follow-up} in the Appendix.
Our sample also contains three sources which were confirmed in later studies to be brown dwarf companions. All followed-up sources are listed in Table~\ref{tab:followedupbds}.

For a single false positive among 30 followed-up sources, it appears our sample has a false positive rate below 5\%.
Fortunately, we do not have to gauge the purity of our sample in a vacuum, as many of the sources in our sample have been the subject of several spectroscopic surveys, with many having been previously confirmed to host brown dwarf companions.
Firstly, \cite{Marcussen2023} aimed to identify false positives for systems consistent with planetary masses, and we retain $3$ of their $9$ exoplanets, while retaining none of their $5$ identified false positives.
\cite{Unger2023} investigated $31$ \gaia orbital solutions with available radial velocities to place better constraints on the companions, of which we retain $7/10$ brown dwarfs, $1/3$ exoplanets, and $0/10$ false positives.
\cite{Stefansson2025} followed up the lowest mass function NSS sources, which had primaries with predominantly low masses. These are the most prone to false positives, and therefore should have the highest rate of false positives. We retain the only brown dwarf in the sample, $5/11$ of the sources labelled as exoplanets, and $8/29$ false positives.
\cite{Barbato2026} used high resolution spectroscopy to update and confirm brown dwarf orbits, and we retain $6/8$ brown dwarfs and $1/6$ false positives.
Finally, \cite{Marcussen2026} obtained high resolution spectroscopy for the brightest remaining sources, from which we retain $7/7$ substellar companions and $0/13$ false positives.
Additionally, our follow-up contained sources later proven to be veritable brown dwarfs: two in \cite{Marcussen2026} and one in \cite{Barbato2026}.

We retained a small sample of planets, a considerable majority of known followed-up brown dwarfs, and a small fraction of the false positives.
In particular, \cite{Barbato2026} and \cite{Marcussen2026} are recent high-resolution spectroscopic follow-up campaigns with samples created similarly to ours, specifically aimed at identifying brown dwarfs in low mass function systems with \gaia orbital solutions, and we have near perfect brown dwarf retention and false positive elimination.
The distribution of $\fmast$ in our sample is shown in Fig.~\ref{fig:massfunction} for both the parent sample and those retained from the spectroscopic cut; and as can be seen, the majority of low $\fmast$ sources are likely false positives, as this cut eliminates approximately $70\%$ of low-mass function sources, which is consistent with the found purity from the follow-up campaigns.
Additionally, the majority of the lowest mass function sources, and therefore the ones easiest to mask as a twin binary, have been followed-up.
Fortunately, now every source with a companion mass estimated below $23\mjup$ has been followed-up, so there is very little uncertainty in the sources at the lowest mass end.
Higher mass sources would be harder to mistake for a false positive, as the corresponding near-twin binaries would have mass ratios further away from unity, thereby making spectroscopic identification of the companion simpler.
This provides good evidence of the high purity of our remaining sample, giving pause to the following statistical results.
For the remainder of our analysis, we assume a uniform contamination rate of 5\% to curb the strength of the results knowing we can not guarantee perfect purity in the sample.
Our sample contains $31$ confirmed brown dwarfs and exoplanets, $4$ spurious orbits (truly single stars) and $7$ false positive near-twin binaries, which is summarised in Table~\ref{tab:confirmedbds}.

After eliminating the false positives, our final sample has 368 binaries.
The distributions of relevant physical parameters (primary mass, secondary mass, period, eccentricity, metallicity, and apparent magnitude) are shown in Fig.~\ref{fig:SampleCorner}.
Fig.~\ref{fig:SampleDiagnostics} shows the confirmed sources and false positives, in addition to detection probabilities for the NSS catalogue (App.~\ref{app:desert}).
The detection probabilities roughly correspond to the source density, granting confidence that the sample is constructed and modelled correctly.

There are a few caveats in the SED vetting worth discussing.
There is a lack of companions near $G=9$, or at primary masses of around $0.7\msun$ (top right panel, Fig.~\ref{fig:SampleDiagnostics}), which is a feature of the SED false positive cut, where the spectral modelling in this mass regime makes it challenging to decisively label a system as a single or a binary \citep{Pennell+2026}.
For this reason, our sample has reduced coverage of these primaries.
Additionally, the false positive rate is likely increased at higher metallicities, since such systems are redder, therefore sit above the main sequence and are more likely to be confused for a twin binary.
There are only 24 systems with a high metallicity $\mathrm{[M/H]}>0.5$, which can be seen in Fig.~\ref{fig:SampleCorner}, and they take on the most likely brown dwarf parameters (high brown dwarf mass, intermediate eccentricity, orbital period near the baseline) and therefore removing them does not meaningfully influence results of the following analyses. 

\section{\gaia selection function inversion} \label{sec:Gaiamock}
\begin{figure*}[!t]
    \centering
    \includegraphics[width=\linewidth]{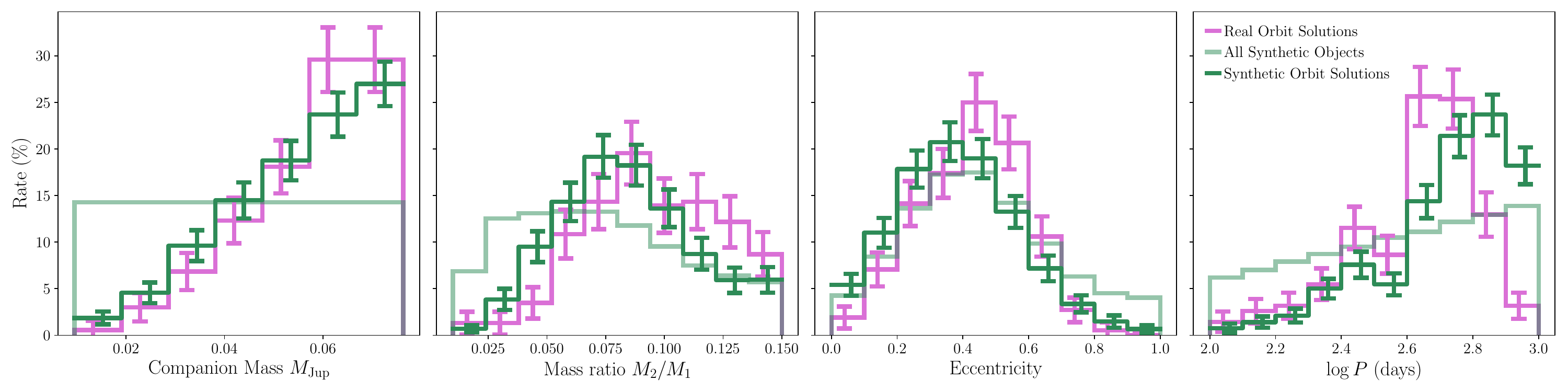}
    \caption{Fiducial models propagated through {\tt Gaiamock} compared to the MS-BD sample, demonstrating that the fiducial models generate good fits to the observed population parameters when selection is accounted for. Bin heights are set to fraction of sources of that sample in the bin such that the areas under each histogram is the same. Error bars on the observed distributions are computed with Poisson uncertainties with an additional 5\% uncertainty to account for remaining spurious solutions (Sect.~\ref{sec:Contamination}). Error bars on the synthetic sample are estimated via bootstrapping: samples are drawn from the whole sample of binaries such that they have the same number of NSS sources $1000$ times, the error bars represent 16th and 84th percentile in each bin. A curious lack of observed sources near the DR3 baseline is currently unexplainable.}
    \label{fig:FiducialComparison}
\end{figure*}
Forward modelling is the primary mechanism which allows the full characterisation and increased understanding of the intrinsic MS-BD population from the observed sample.
\texttt{Gaiamock} \citep{ElBadry+2024} is a generative forward model for the \gaia DR3's astrometric NSS pipeline that reproduces the scanning cadence of the telescope and uses realistic uncertainties to generate mock astrometric measurements, and then fits astrometric models and assigns full orbit solutions using the published selection criteria. 

As a first test, we compared the distribution of orbital parameters in the MS-BD binaries to those of stellar binaries.
In general, we can choose some distribution of the relevant orbital parameters (the companion mass $M_2$, orbital period $P$, and eccentricity $e$), and marginalise over the remaining nuisance parameters such as the orientation of the orbit.
To start, we constructed a representative parent sample (Appendix~\ref{app:parentsample}) by identifying all sources which would have entered the DR3 astrometric cascade after preprocessing \citep{Riello2021, Halbwachs+2023}, in addition to passing the cuts used in creating our MS-BD sample such that we retain a sample which contains all sources which could have, in principle, ended up in our MS-BD sample.
One can create a synthetic dataset by sampling from this parent distribution, assigning companions based on the chosen models, and forward model whether such a system would enter \tgaia's NSS catalogue.
This effectively marginalises over the distribution of primary masses, distances, and sky position, alleviating the need to impose such distributions, while also accounting for any biases introduced in creating our MS-BD sample.

Surveys of solar type stars find a log-normal distribution in orbital periods peaking at $10^5$\,days, with a spread around 2\,dex, and a mass ratio distribution that is largely uniform aside from an excess of twins \citep{Raghavan+2010, DucheneKraus2013, MoeDiStefano2017}. 
Additionally, \gaia has revealed that at intermediate separations, where it assigns orbit solutions (${\sim}\!10^2$--$10^3$ days), the eccentricity distribution of lower main sequence primaries with stellar companions is similar to a normal distribution peaking at $\bar{e}=0.4$ with a spread of $\sigma_e=0.2$ \citep{Wu+2025, Pennell+2026}.
We simulated $150~000$ binaries following these fiducial models, determining which would be assigned an orbit solution from the \gaia DR3 astrometric pipeline.

The resulting synthetic orbit solutions are compared to the MS-BD sample in Fig.~\ref{fig:FiducialComparison}, where the distributions of primary mass, mass ratio, and eccentricity are shown for both populations.
We see broad agreement between brown dwarfs and stars as companions in the distributions of all these parameters.
The selection function is relatively straightforward averaged over the population: high masses and low eccentricities are preferred, with a more complicated dependence on the period due to degeneracy with parallax and its harmonics.
For this reason it is possible to fairly easily compare subsamples directly to see if they are qualitatively different from each other, since we would not expect the selection function to vary significantly between them.
We therefore show the distributions and covariances of relevant parameters -- primary mass, secondary mass, mass ratio, period, eccentricity, metallicity, and apparent magnitude -- in (Fig.~\ref{fig:SampleCorner}).
The main effect of changing the mass of the companion is that there is a higher representation of more extreme values of orbital period, eccentricity and metallicity.
Higher companion masses create larger astrometric wobbles and therefore enter our sample more often Fig.~\ref{fig:FiducialComparison}, and the period, eccentricity and metallicity values only seen in the high mass companions are precisely those which are intrinsically less common, making this consistent with selection effects.
We see consistency with a uniform companion mass and mass ratio distribution, which is further explored quantitatively in App.~\ref{app:desert}, where we show there is no dearth of companions at intermediate masses as suggested by \cite{MaGe14, Stevenson+23}.
One caveat is the apparent lack of observed systems near \tgaia's DR3 baseline, which also seen from the detection curves in Fig.~\ref{fig:SampleDiagnostics}.
This, does not appear to be real, or the result of the fiducial model choice, as both planetary and stellar companion orbital period distributions peak above this range \citep{Raghavan+2010, Bryan2016, MoeDiStefano2017}, making it more likely that this is a by-product of the sample creation or forward modelling.

\section{Eccentricity distribution} \label{sec:Eccentricities}
From visual inspection of Fig.~\ref{fig:FiducialComparison}, the eccentricity distribution appears to be similar to a normal distribution centred around $e=0.45$, which is consistent with general surveys of stellar binaries in the field \citep[e.g.][]{Duquennoy1991}.
In particular, the eccentricity distribution of selection-corrected \gaia orbit solutions follows a normal-like distribution with an average eccentricity $\bar{e}=0.4$ for M- to G-type primaries \citep{Wu+2025, Pennell+2026}.
Conversely, the eccentricity distribution for large planets at these separations is far lower, near $\bar{e}=0.2$ \citep{Bowler2020, Bitsch2020}.
Planets generally have low eccentricities because the damping timescale of the eccentricity in a protoplanetary disk is small \citep{Goldreich1980}, which would therefore require interactions after the disk evaporation to pump up a planet's eccentricity, such as the scattering of another planet \citep[e.g.][]{Ford2008}.
Therefore, inflated eccentricities speak to the brown dwarfs in our sample not being born in protoplanetary disks.

To more quantitatively describe and interpret the eccentricity distribution, we constrain a parametrised model.
Computing the likelihood of a particular binary given a parametrised model contains three parts: a model $M(e\mid\vec{\theta})$ with a set of flexible set of parameters $\vec{\theta}$, the selection function $S_\mathrm{NSS}(e)$ which determines the probability of being assigned an orbit solution, and the observed parameters with their uncertainties. 
The observed eccentricity distribution $\prob{e}{\vec{\theta}}$ corresponds to the intrinsic distribution $M$ propagated through the selection function $S_\mathrm{NSS}$,
\begin{equation}
    \prob{e}{\vec{\theta}} = \frac{1}{Z}\int M(e\mid\vec{\theta}) S_\mathrm{NSS}(e)~\di e,
\end{equation}
where $Z$ is simply a normalisation constant.

For an observed binary $i$ with a best fit eccentricity $e_i$ and Gaussian uncertainty $\sigma_{e,i}$, the likelihood of a parametrised model to reproduce the observed eccentricity is
\begin{equation}
    \mathcal{L}_i = \ln \mathcal{L}(i\mid {\vec{\theta}}) = \ln\int\mathrm{exp}\biggl(-\frac{(e-e_i)^2}{\sigma_{e,i}^2}\biggr) \prob{e}{\vec{\theta}}~\di e + c, \label{eq:likelihood}
\end{equation}
where the constant $c$ comes from normalisation of the likelihood which become an additive constant in the log-likelihood, which therefore makes it irrelevant for constraining the parameters.
The log-likelihood of the whole sample is simply the sum of the log-likelihood for each source.
This likelihood can be computed numerically straightforwardly by constructing each constituent probability as a probability grid, and multiplying them together element-wise and summing the result, which in the limit of a finer grid is the integral described above.

The probability $S_\mathrm{NSS}(e)$ was computed object-wise. For each source in our sample, we simulated 500 binaries with random orbital angles and distances to estimate the probability of such a binary of entering the NSS catalogue marginalised over these parameters.
As for the parametrised model, we used a Gaussian eccentricity distribution parametrised by a mean $\bar{e}_0$ and spread $\sigma_0$ since the eccentricity distribution is well described by a Gaussian at different companion masses and periods (Fig.~\ref{fig:SampleCorner}).
To consider how the eccentricity changes with orbital period and brown dwarf mass, we defined the peak of eccentricity as a power law with its pivot points in mass an period as additional parameters,
\begin{equation}
    \bar{e}(M_2,P) = \bar{e}_0 \biggl(\frac{M_2}{42.5\mjup}\biggr)^\alpha \biggl(\frac{P}{1~\text{year}}\biggr)^\beta.
\end{equation}
We set the mass pivot point at $42.5\mjup$, which is near the centre of the companion mass parameter space, and also where \cite{MaGe14} noted distinct populations in their brown dwarf sample. This provides a test to see if our sample is consistent with two separate populations.
We set the pivot point at a year for being near the centre of the period parameter space logarithmically. 
For this model, the set of parameters is therefore $\vec{\theta} = (e_0,\sigma_0,\alpha,\beta)$.

Finally, we introduced a mixture term to account for the false positives $\mathcal{L}_\text{err}$.
The false positive term $\mathcal{L}_\text{err}$ was computed identically to $\mathcal{L}_i$, except the eccentricity model is that of stellar binaries from \cite{Wu+2025} where $\bar{e}\approx0.4$ and $\sigma_0\approx0.2$ because the false positives are likely to be such stellar companions. 
This choice serves only to increase the significance of the measured $\alpha$ and $\beta$, as the false positive term biases $\alpha$ and $\beta$ towards zero since it is intentionally agnostic to them.
Furthermore, this effectively prevents any individual measurements from contributing significantly to the likelihood since $\mathcal{L}_\text{err}$ sets a minimum likelihood per-source.
Since our sample has an apparent contamination below 5\%, we made the false positive rate a parameter $F_p$ in the mixture model, which is allowed to vary between 0 and 5\%, making the full likelihood
\begin{equation}
    \mathcal{L}_\text{tot} = (1-F_p)\mathcal{L}_i + F_p\mathcal{L}_\text{err}.
\end{equation}

We sampled the posterior distribution of this likelihood using the affine-invariant ensemble Markov chain Monte Carlo (MCMC) sampler implemented in the Python package `emcee` \citep{Foreman-Mackey2013} for 100\,000 iterations with 25\,000 iterations of burn-in, as all the parameters upon visual inspection appear converged after only a few thousand.
The results are shown in Fig.~\ref{fig:EccentricityFit}, with mild evidence that the mass-dependence $\alpha=-0.13\pm0.09$ is negative, and that the period-dependence $\beta=0.10\pm0.05$ is positive.
The false positive rate is poorly constrained, meaning these results are robust to whatever modest contamination rate exists, and are not being influenced by a few outliers.

\begin{figure*}[!t]
    \centering
    \includegraphics[width=\linewidth]{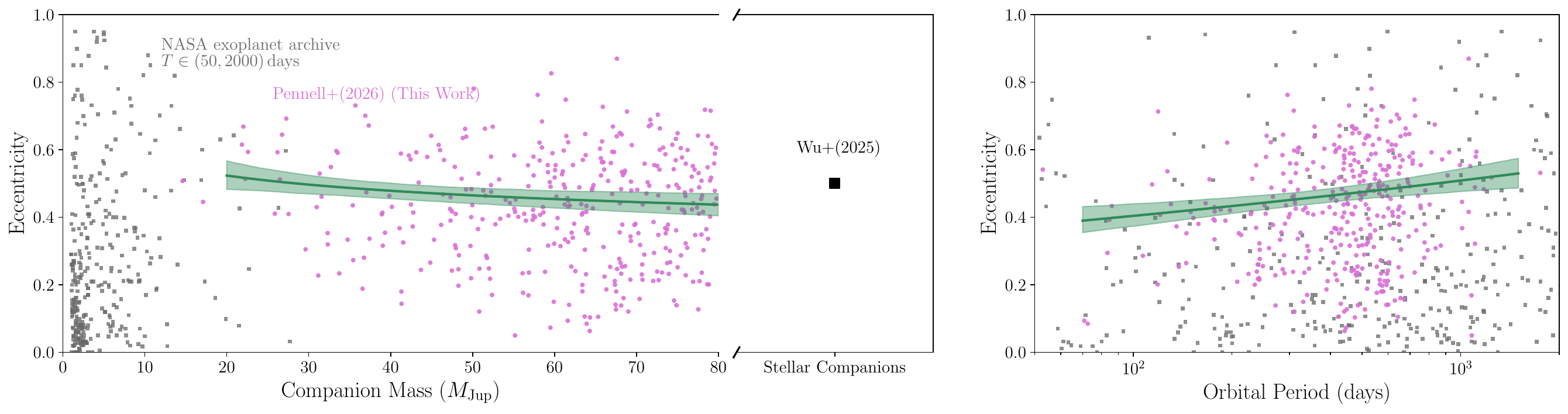}
    \caption{Results for the mass dependence on the mean eccentricity for substellar companions compared with other observed samples. The individual eccentricities of our MS-BD sample is shown in pink. Our best fit results from Fig.~\ref{fig:EccentricityFit} are shown in a green line,  with the shaded regions representing the 16th and 84th percentile of 1000 sets of parameters sampled from the computed posteriors. 
    We compare this to the mean eccentricity for stellar companions to solar type stars with orbit solutions in \tgaia \citep{Wu+2025}. Our results show broad agreement, with limited evidence for a decreasing mean eccentricity as a function of mass. We also compare to planetary eccentricities for large planets from the NASA exoplanet archive \citep{Christiansen2025} in grey.}
    \label{fig:EccentricityResults}
\end{figure*}

We compare with planetary and stellar companions in Fig.~\ref{fig:EccentricityResults}, where we also show our entire MS-BD sample in comparison with all 327 planets with periods between 50 and 2000\,days and masses or minimum masses over the mass of Jupiter from the NASA exoplanet archive \citep{Christiansen2025}.
This is useful point of comparison, taking the planets which coincide with our astrometrically sensitive period range, as it shows that the eccentricities are generally lower in the planets than even our lowest-mass brown dwarfs.
This is also compared to \cite{Wu+2025}, who assembled a sample of solar type stars with stellar companions in \tgaia's NSS catalogue and report a near-Gaussian (Rayleigh) distribution with a mean eccentricity around $0.4$.
We show a moderately decreasing mean eccentricity with companion mass which is consistent with both results, but is also broadly consistent with being flat.

While the sample is sparse at low brown dwarf masses $\lesssim20\mjup$, all such brown dwarfs in the sample have high eccentricities, while the \gaia selection function is biased towards low eccentricities (Fig.\ref{fig:FiducialComparison}), while the planets up to ${\sim}10\mjup$ generally have lower eccentricities.
For this reason, the \gaia DR3 sample shows no clear physical transition of eccentricities at any brown dwarf mass.

\section{Metallicities} \label{sec:Metallicities}
An additional upside of constructing a sample where companions do not contribute significantly to the light is that their spectra can more precisely yield masses and metallicities since we have removed the possibility of a contaminating companion, and we assume the brown dwarf metallicity is that of its primary \citep{El-Badry2018}.
Metallicities are computed by \cite{Li+2025} through creating an emulator for the XP spectra and as a result the measurement uncertainty and also the statistical power of a subsample depends on the coverage of parameter space of the training sample.
At near solar metallicity, the metallicity measurement uncertainty is less than 0.1\,dex, but this rises where there are fewer sources and for this reason the uncertainty rises at extreme metallicities.

\begin{figure}
    \centering
    \includegraphics[width=\linewidth]{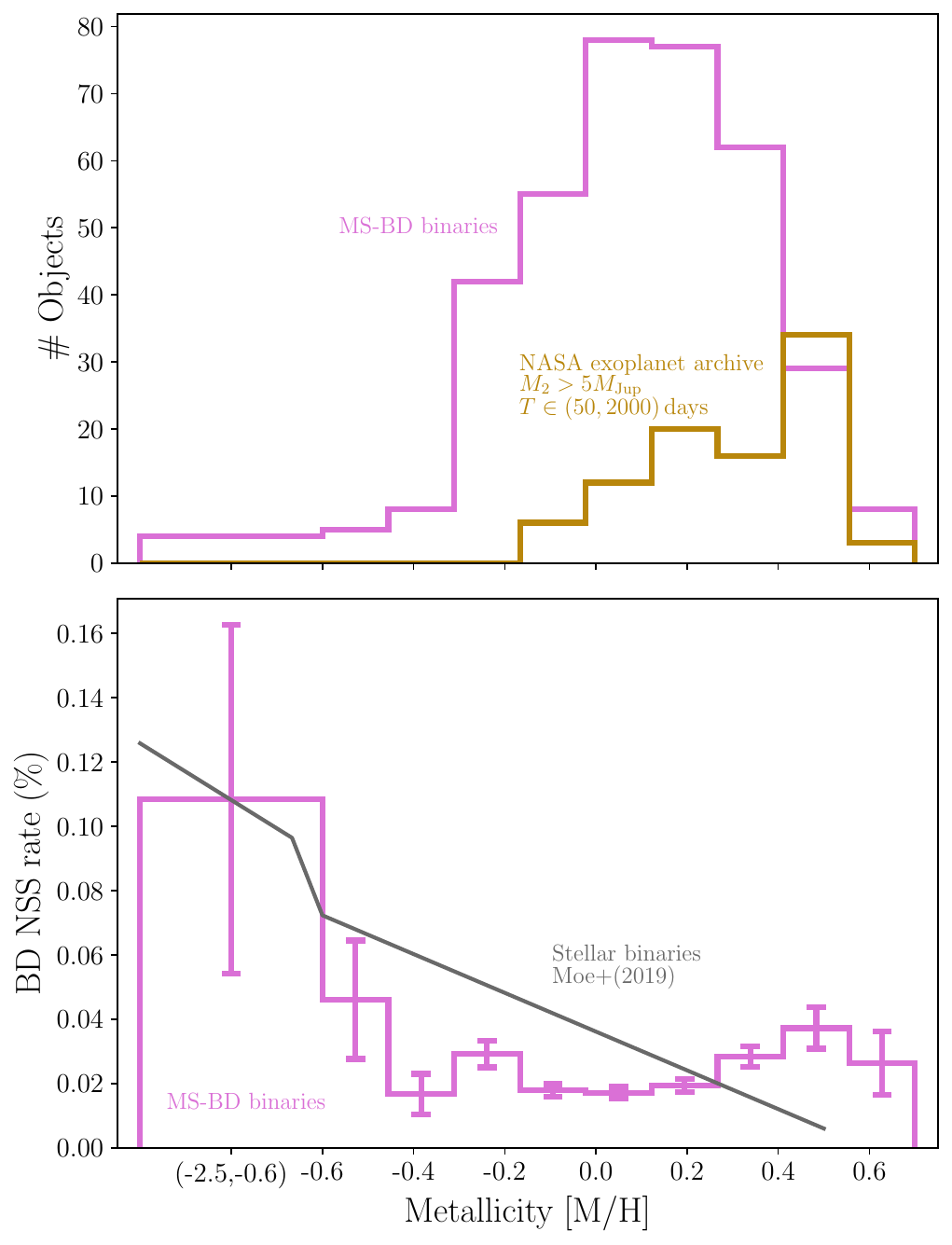}
    \caption{Distribution of metallicities [M/H] of our MS-BD systems compared to stellar and planetary companions. Bottom panel: distribution of metallicities in the MS-BD sample (pink) compared to that of all large planets ${>}5\mjup$ at astrometrically detectable periods $P\in(50,2000)$\,days from the NASA exoplanet archive \citep[gold,][]{Christiansen2025}. This demonstrates the consistency of MS-BD binaries with stellar binaries, and inconsistency with planetary systems. Caution should be taken at the extreme metallicities, as these measurements are more uncertain and the errorbars represent lower bounds. Top panel: rate of MS-BD binaries in bins of metallicity by comparing the MS-BD sample to the representative parent sample (App.~\ref{app:parentsample}) shown in pink, compared to the rate of binarity for stellar binaries from \cite{Moe2019}, where the y-axis is meaningless.}
    \label{fig:metallicities}
\end{figure}

In the top panel of Fig.~\ref{fig:metallicities}, we show the metallicity distribution of all currently known host stars with planets of masses greater than that of $5\mjup$ with orbital periods approximately in the range of \tgaia's astrometric sensitivity $P\in(50,2000)$\,days from the NASA Exoplanet Archive \citep{Christiansen2025}.
The planetary systems show a strong preference for higher metallicities, with a complete dearth of systems at $\mathrm{[M/H]}<-0.2$, 
while our sample contains brown dwarf companions at preferably lower metallicities.

We compare the metallicity distribution of our MS-BD sample to the representative parent population sample (App.~\ref{app:parentsample}) to compute the occurrence rate as a function of metallicity, which is shown in the bottom panel of Fig.~\ref{fig:metallicities}, where the multiplicity rate of stellar binaries with solar-type primaries at close ($<10$\,AU) separations is strongly anticorrelated with metallicity \citep{Raghavan+2010, Moe2019, El-Badry2019}.
We can reasonably extrapolate to the total brown dwarf population from these comparisons with little concern for the selection function because astrometry is relatively agnostic to the metallicity of the host star.
The trend in multiplicity from \cite{Moe2019} is compared to our MS-BD sample (Fig.~\ref{fig:metallicities}), where we see a rise of companions at low metallicities as expected, however we our sample is instead roughly flat between $\mathrm{[M/H]}=-0.4$ and $0.65$.

The uncertainties shown are the Poisson uncertainty corresponding to the number of MS-BD binaries in each bin. 
However, this should only represent a lower limit on the true uncertainty since this does not incorporate the (unknown) uncertainties from the poorer performance.
The uncertainties certainly increase at the extreme metallicities as previously described, but are also dependant on how the sample is constructed, since preprocessing in \gaia depends on the astrometric and photometric measurements \citep{Halbwachs+2023} which may interact differently for the binaries and the truly single stars in the parent sample.

\section{Discussion} \label{sec:discussion}
We find the eccentricity and metallicity distributions of the MS-BD sample are inconsistent with the suite of massive planets from the NASA exoplanet archive \citep{Christiansen2025} even for the lowest brown dwarf masses, which implies these brown dwarfs are distinct from their planetary counterparts.
Comparing to this large, heterogenous catalogue is naturally not a perfect point of comparison, as these are generally not found astrometrically and there is no straightforward way to model the selection or biases in the sample, and for this reason we make no quantitative comparisons.
Furthermore, metallicities the planetary metallicities have been assigned through a variety of schemes, and many were computed as iron metallicities $\mathrm{[Fe/H]}$, therefore being slightly offset from the computed global metallicity $\mathrm{[M/H]}$ of the MS-BD sample.
Regardless, this remains a useful qualitative comparison as the selection on eccentricity and metallicity of our sample is generally relatively straightforward, and the difference between the MS-BD sample and these planets is sufficiently extreme that it is challenging for selection characteristics to reconcile the differences.

We find high eccentricities broadly consistent with values found for stellar companions, well fit by Gaussian distributions peaking at intermediate eccentricities for all brown dwarf masses and across all periods we are sensitive to \citep[e.g.][]{Duquennoy1991, Raghavan+2010, DucheneKraus2013, MoeDiStefano2017}.
While the brown dwarf companions appear to behave like the stellar binaries discussed by \cite{Wu+2025}, where moderate mean eccentricities can be explained by several different kinds of physics, and they remain cautious as to what physics is generating these eccentricities.
For example, circumbinary disk interactions can settle into intermediate eccentricities, estimated to be near $e=0.45$ by \cite{Lai2023}. Additionally, simulations of scattering of planets or brown dwarfs can excite the eccentricity of the remaining components to these intermediate levels, with an apparent anticorrelation of eccentricity with companion mass, which we see mild evidence for \citep{Chatterjee2008, Ford2008, Carrera2019}.
As a result, while the eccentricities are in disagreement with planets, it remains challenging to identify the dominating dynamical processing mechanism.

We see a rise in the rates of brown dwarf companions at low metallicity, which is found at these separations for stellar binaries \citep{Moe2019}.
However, previous studies show a consistent decrease in multiplicity with increasing metallicity, while our sample shows a relatively constant rate above $\mathrm{[M/H]}>-0.4$, with an apparent rise to supersolar metallicities.
This rise does not appear to correspond with a ``planet-like" channel, as these sources are predominantly the highest mass companions (Fig.~\ref{fig:SampleCorner}).
Instead, this may be partly due to false positives, since at high metallicities sources are redder and sit furthest off the main sequence, and will therefore have spectral shapes more similar to the twin binaries that occupy the same region of parameter space.
This makes them harder to discern, and therefore increases the false positive rate.
This points to a general limitation of the SED identification scheme, which will be at least partially alleviated in the imminent fourth data release from \tgaia, where published epoch measurements and broader publication of the high resolution RVS spectra \citep{Recio-Blanco2023} will allow for better identification of false positives and stellar properties.

In addition to methodological limitations, the most comparable metallicity-multiplicity studies \citep[e.g.][]{Moe2019} were focused on solar-type primaries, while the majority of our sample have M-type (${\lesssim}0.65\msun$) primaries.
This makes the comparison challenging, as lower mass stars are broadly observed to have different multiplicity characteristics compared to solar type stars such as lower orbital periods and higher mass ratios \citep{DucheneKraus2013, Bate2019, Winters2019}, which may correspond to changes in formation also influenced by metallicity.
Specifically, companions at these separations are likely to have formed through instabilities in the protostellar disk, which depends on the relative mass of the disk to the primary and the thermal conditions of the disk, which may differ by host and environment \citep{Kratter2016}.
In massive stars, for example, the high multiplicity at close separations \citep{Sana2011} is explainable with their generally high relative disk masses \citep{Oliva2020}.
Generally, the relative mass of the disk to the primary is fairly constant for low mass stars \citep{Andrews2013}, and since the formation depends on mass ratio, not companion mass \citep{Kratter2016}, varying primary masses would change not just the thermal properties of the disk but also the expected companion masses formed.
It warrants further investigation of protostellar disks around stars of different masses to determine how the radiation structure and mass budget may be influenced by metallicity.

For the MS-BD sample, the eccentricities and metallicities remain comparable to stellar binaries down to the lowest masses, and additionally the companion masses, mass ratios, and orbital periods imply no variation in the population across any of the parameters.
This paints an interesting picture, as previous studies have found distinct populations by brown dwarf mass which is also revealed in eccentricity \citep{GretnerLineweaver06, MaGe14} which pointed to where core-accretion ``planet-like`` formation and instability ``star-like`` formation dominate.
Conversely, our sample has no such signature, which therefore suggests that if there is such an observable turnover, it happens near or below the brown dwarf mass limit.
From a visual inspection of Fig.~\ref{fig:EccentricityResults} giant planets up to ${\sim}10\mjup$ remain at low eccentricities, implying a rather narrow turnover from core-accretion to disk-instability formation between ${\sim}10$ and ${\sim}20\mjup$. 
This is consistent with lower limits for star formation in simulations, where \cite{Bate2012} simulates star formation resulting in masses down to ${\sim}15\mjup$, arguing that while collapse may happen at masses of only a few Jupiter masses, subsequent accretion raises the minimum likely mass after formation has ceased. This is further shown to have a similar boundary at different metallicities \citep{Bate2019}.

\section{Summary} \label{sec:Conclusion}
We assembled a well-understood sample of 368 main sequence primaries with brown dwarf companions from \gaia DR3's published astrometric solutions, after using Gaia XP spectra to eliminate ${\sim}70\%$ of the initial candidates as being likely near-twin false positives. The resulting sample is quite pure: we retain the vast majority of confirmed brown dwarfs and very few false positives, verified via several external high-resolution spectroscopic surveys constructed in the same way as our sample \citep{Barbato2026, Marcussen2026}.
This again demonstrates the necessity and utility of cleaning out false positives, via spectroscopy or SEDs, from astrometric samples where the companions are expected to be near-dark.
We studied the population properties of the brown dwarf companions through modelling the selection function from \gaia using the generative forward model {\tt Gaiamock}. Primarily, we found that brown dwarf companions resemble low-mass stellar companions with respect to their eccentricity and metallicity distributions, (Fig.~\ref{fig:FiducialComparison}), independent of their masses or orbital periods (Fig.~\ref{fig:SampleCorner}).

We showed that eccentricities have an average value similar to stellar binaries in \tgaia's astrometric orbit catalogue, while being higher than planetary eccentricities (Fig.~\ref{fig:EccentricityResults}).
The eccentricity distribution barely depends on mass ($\alpha=-0.13\pm0.09$), for all $M>20M_{Jup}$, or on period ($\beta=0.10\pm0.05$).
Finally, we found that the metallicity distribution of stars with brown dwarf companions is dramatically inconsistent with that of stars hosting planets, which are predominantly found at higher metallicities (Fig.~\ref{fig:metallicities}). The incidence of brown dwarf companions shares the same anti-correlation found in stellar binaries \citep{Moe2019}, with many of our sources being at low metallicities.
In the appendix, we further demonstrated how our sample provides no significant evidence for a ``desert'' at intermediate brown dwarf masses ($35$--$55\mjup$ from \cite{MaGe14}, in the period range we are sensitive to Fig.~\ref{fig:DesertLikelihood}). There appears to be no dearth of companions at any mass or period regime in our sample.
This implies that there is a homogenous population across all probed brown dwarfs. Therefore, if there is a turnover between ``planet-like'' and ``star-like'' population, it appears to occur below $20\mjup$.

We can not directly see how stars, brown dwarfs, or planets form.
And while it is not possible to pin a particular eccentricity or metallicity on a particular formation mechanism, the distributions compared to those of stellar and planetary companions are revealing.
But fortunately, well understood samples of ever-increasing sizes are becoming readily available alongside improving theoretical predictions and simulations.
The imminent release of \gaia DR4 will produce a larger sample of binaries extending to higher periods, with a better ability to model outliers and false positive through the published epoch observations.
In our sample, the brown dwarf companions across all masses have metallicities and eccentricities rather consistent with stellar binaries, and wholly inconsistent with planetary companions.
And so, to this end, it seems that a brown dwarf as a companion is dynamically ``just a star''.

\begin{acknowledgements}
B.~Pennell, H.~W.~Rix, S.~Shahaf, J.~M\"uller-Horn, and J.~Li acknowledge support from the European Research Council for the ERC Advanced Grant [101054731]. 
This work has made use of data from the European Space Agency (ESA) mission \textit{Gaia} \url{https://www.cosmos.esa.int/gaia}, processed by the \textit{Gaia} Data Processing and Analysis Consortium (DPAC, \url{https://www.cosmos.esa.int/web/gaia/dpac/consortium}). Funding for the DPAC has been provided by national institutions, in particular the institutions participating in the \textit{Gaia} Multilateral Agreement.
The authors are grateful to Angela Hempel, Maren Hempel, Simón \'Angel, and Paul Eigenthaler for their help and support with observations at the ESO/MPG 2.2m telescope. Absolutely no generative AI was used in the writing of this text.
\end{acknowledgements}
\bibliographystyle{bibtex/aa} 
\bibliography{references.bib}

\begin{appendix}
\nolinenumbers
\section{Creating a parent sample} \label{app:parentsample}
\begin{figure}[!h]
    \centering
    \includegraphics[width=\linewidth]{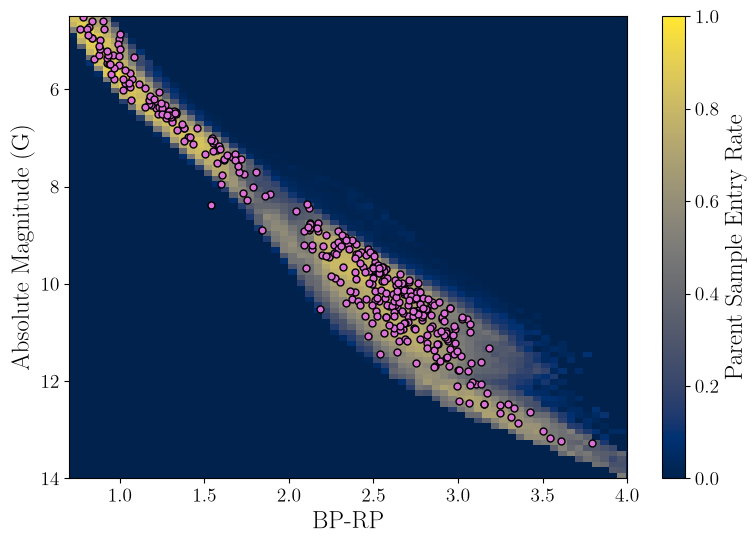}
    \caption{Rate for any given source (Right panel, grey dots, Fig.~\ref{fig:SampleDiagnostics}) to enter into the representative parent sample as a function of \gaia colour and absolute magnitude. We find good agreement with the distribution of sources in the parent sample across the CMD as compared to the distribution of the MS-BD sample, which gives pause.}
    \label{fig:SampleSelection}
\end{figure}

To make a direct comparison between a synthetic and real sample of orbit solutions, we assembled the ``parent" population of primary stars, and sampled from them, assigning arbitrary companions to create a synthetic sample of binaries which match the real sample in terms of distance, primary mass, and apparent magnitude, such that these do not need to be estimated.
As a result, only the choice of fiducial models for the orbital parameters of the companions can contribute to the distributions in Fig.~\ref{fig:FiducialComparison}.

We first followed the sample creation for the MS-BD sample in Sect.~\ref{sec:Sample}, but not limited to orbital solutions.
We then followed the preprocessing steps outlined in \cite{Halbwachs+2023, Riello2021}, limiting to sources with sufficient observations, and making several cuts to limit sources which may have astrometry affected by a nearby background source or wide companion.
One such cut involved comparing the G- and BP/RP-bands of the \gaia observations, since the window sizes are different, and their inconsistency could indicate the presence of a contaminating source \cite{Riello2021}.
As pointed out in \cite{Riello2021}, this photometric cut cleanly cuts the main sequence at the bottom.
However, we retain sources which lie far above the main sequence, none of which enter the NSS catalogue. 
These are likely to be twin binaries, or young stars, either way unmodellable by the spectroscopic analysis of \cite{Li+2025} or the forward model \cite{ElBadry+2024}.
We restrict our sample to the main sequence with isochrones.
We bound the main sequence from above with a low age $10^{7.4}$\,yr and higher metallicity $\mathrm{[M/H]}=0.3$, and from below with an age of $10^10$\,yr and metallicity of $\mathrm{[M/H]}=-0.6$ based on PARSEC isochrones \citep{PARSEC}, which eliminates less than a percent of the sample.
The result is a set of sources that would be propagated through \tgaia's astrometric pipeline if they had a brown dwarf companion providing a significant astrometric wobble.
This allows for direct computation of detection probabilities in the MS-BD sample.

We show the rate of a source passing all the cuts as a function of colour-magnitude position in Fig.~\ref{fig:SampleSelection}.
As can be seen, the regions where we retain sources coincide with higher density of BD companions, implying that this sample creation is indeed a fairly representative parent sample.

\section{Brown dwarf mass distribution} \label{app:desert}
\begin{figure*}[t]
    \centering
    \includegraphics[width=\linewidth]{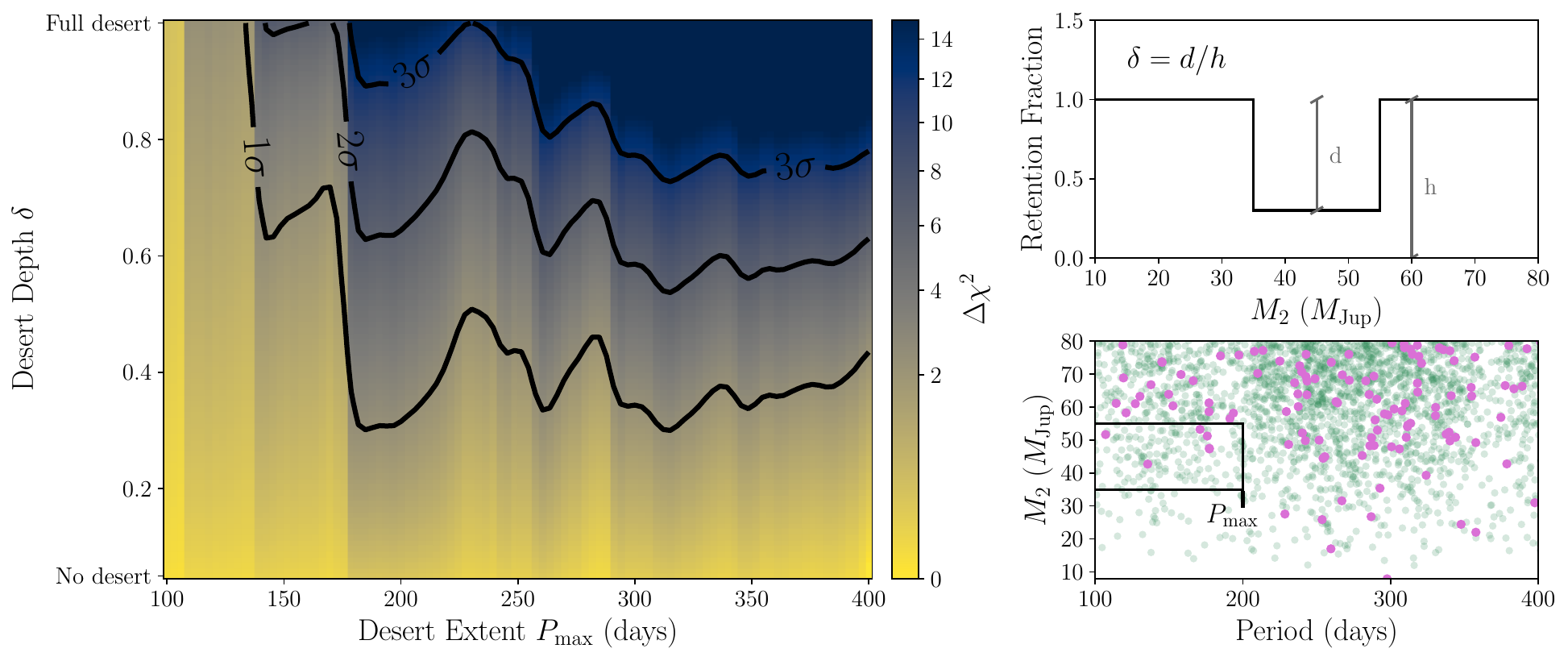}
    \caption{Results of the scheme (Sect.~\ref{sec:Eccentricities}) to quantitatively determine to what degree a brown dwarf desert is excluded by our sample. The right panels schematically show the scheme, where we introduce a $35$--$55\mjup$ gap into the fiducial models outlined in Sect.~\ref{sec:Gaiamock}, parametrised by its maximum period extent $P_\text{max}$ and depth $\delta$. The bottom right panel shows the observed sample in pink, and a random subset of synthetic binaries in green to highlight where our resolving power lies, and why we have relatively little constraining power below 200 days. The left panel shows the $\Delta\chi^2$ for a given set of $(P_\text{max},\delta)$ compared to no desert, demonstrating that no desert is needed to explain the resulting sample.}
    \label{fig:DesertLikelihood}
\end{figure*}
Brown Dwarfs, not dissimilar to planets in the history of modern astrophysics, were first observed in the same year as the first planets \citep{Oppenheimer1995, Nakajima1995, Mayor1995}, and have been the targets of observations and specialised surveys since.
One feature of the occurrence rates of brown dwarfs is the so-called ``Brown Dwarf Desert", which is a term that occasionally causes confusion.
This is because a brown dwarf desert first appeared in the early stages of substellar companion discoveries with truly planetary mass companions generally outnumbering their Brown Dwarf counterparts \citep{Marcy1998, Marcy2000, Halbwachs2000}, yet more recently has been described more specifically as a lack of `intermediate-mass' ${\sim}35$--$55\mjup$ Brown Dwarfs \citep{GretnerLineweaver06, MaGe14, Feng2022}, which may only manifest within orbital periods of a few dozen days \citep{Shahaf2019, Troup2016}, or a hundred days \citep{Stevenson+23}, or several hundreds of days \citep[several AU;][]{Dieterich2012, Chen2024, An2025}, or instead at hundreds of AU \citep{Allen2007, McCarthy2004}, or perhaps there's no desert at hundreds of AU \citep{Metchev2009}, or perhaps that there are simply fewer Brown Dwarf companions of all masses \citep{Kraus2008, Kiefer2019}.

This definition has been hard to pin down due to different surveys being sensitive to different regions of parameter space, such as, for example, the spectroscopic surveys sensitive lower orbital periods such as with APOGEE \citep[e.g.][]{Shahaf2019}, and imaging surveys sensitive to wide binaries \citep[e.g.][]{Allen2007}.
All together, selection effects of these studies are hard or entirely impossible to quantify.
A ``desert" is of course never entirely ``dry'', with intermediate mass BD companions being found \citep[e.g][]{Metchev2004}, with some found even at the closest separations through transits \citep[e.g.][]{Carmichael2021, Zhang2026}, making every claimed desert have numerous ``oases''.

\gaia DR3 astrometry behaves best between 100 and 1000 days (Fig.~\ref{fig:SampleDiagnostics}), meaning we can only probe whether a desert reaches into these periods.
\cite{MaGe14} and \cite{Stevenson+23} Observe a desert for intermediate $(35\mjup$--$55\mjup)$ BD masses within around 100 day orbital periods.
We can directly measure if this desert extends into our sample.
From visual inspection of Figs.~\ref{fig:SampleDiagnostics} and \ref{fig:FiducialComparison}, it appears that there is no evidence for a desert as the brown dwarfs are consistent with a uniform mass distribution.
Furthermore, astrometric summary statistics from {\it Gaia} DR3 have already been found to be consistent with no brown dwarf desert \citep{Wallace2026}.
We can, however, relatively painlessly quantitatively determine to what degree our sample is or is not consistent with such a desert, through a scheme similar to that of Sect.~\ref{sec:Eccentricities}.

A desert would change the inherent distribution of the period $P$ and companion masses $M_2$, $D=\prob{P,M_2}{\text{desert}}$.
This would therefore change the probability of observing a particular period $P_i$ and companion mass $M_{2,i}$ for a source $i$ in our sample, $\prob{P_i,M_{2,i}}{D}$.
We can therefore quantify to what degree a ``desert" model explains our sample compared to fiducial models by computing the likelihood $\likelihood{\{P_i,M_{2,i}\}}{D}$, where $\{P_i,M_{2,i}\}$ represents the set of observed periods and companion masses and their uncertainties.
The likelihood is identical to that of Eq.~\ref{eq:likelihood}, except with two parameters, with a Gaussian uncertainty contribution for each.

We can now choose how to parametrise the desert.
We maintained the fiducial period and companion mass distribution from Fig.~\ref{fig:FiducialComparison}, but introduced a parametrised ``notch" to represent the desert found by \cite{MaGe14}, between $35$ and $55\mjup$ extending to some maximum period $P_\text{max}$ and depth $\delta$, which is shown schematically on the right panels of Fig.~\ref{fig:DesertLikelihood}.
This allows for direct comparison with the fiducial model where $P_\text{max}=100$ and $\delta=0$.

Finally, we attempted to estimate \tgaia's selection function $\prob{\text{NSS}}{P_i, M_{2,i}}$.
We simulated $1\,000\,000$ binaries from the same scheme as in Sect.~\ref{sec:Gaiamock} with companions distributed uniformly between $10$ and $80\mjup$, and logarithmically uniform between $10$ and $10^4$\,days.
We placed these synthetic binaries into 200 bins of companion mass and 200 bins of logarithmic period.
We determined the rate of getting an orbit solution in each bin, which represents the probability for a source with a companion of a particular orbital period and companion mass to be assigned an orbit solution.
This probability is naturally marginalised over the primary mass, apparent magnitude, and distance distribution of the parent sample (App.~\ref{app:parentsample}), in addition to the choice of eccentricity distribution and the orbital angles.
We applied a broad smoothing filter to estimate the detection curves in Fig.~\ref{fig:SampleDiagnostics}.

Finally, we allowed for a 5\% contamination fraction. We do this by creating a mixture model through adding a contamination term $\mathcal{L}_\text{err}$ in addition to the $\mathcal{L}_i$ just as in Sect.~\ref{sec:Eccentricities}
\begin{equation}
    \mathcal{L}_\text{tot} = 0.95\mathcal{L}_i + 0.05\mathcal{L}_\text{err},
\end{equation}
where $\mathcal{L}_\text{err}$ is computed identically to $\mathcal{L}_i$, except with the fiducial models with no notch.
This represents the worst-case scenario as this will simply reduce the likelihood difference compared to the null hypothesis, and effectively prevents any individual measurement from contributing significantly negatively to the likelihood since $\mathcal{L}_\text{err}$ is a constant.
We do this to strengthen the significance of the reported results.

\begin{figure*}[!t]
    \centering

    \begin{minipage}[t]{0.48\textwidth}
    \vspace{0pt}
        \centering
        \includegraphics[width=\linewidth]{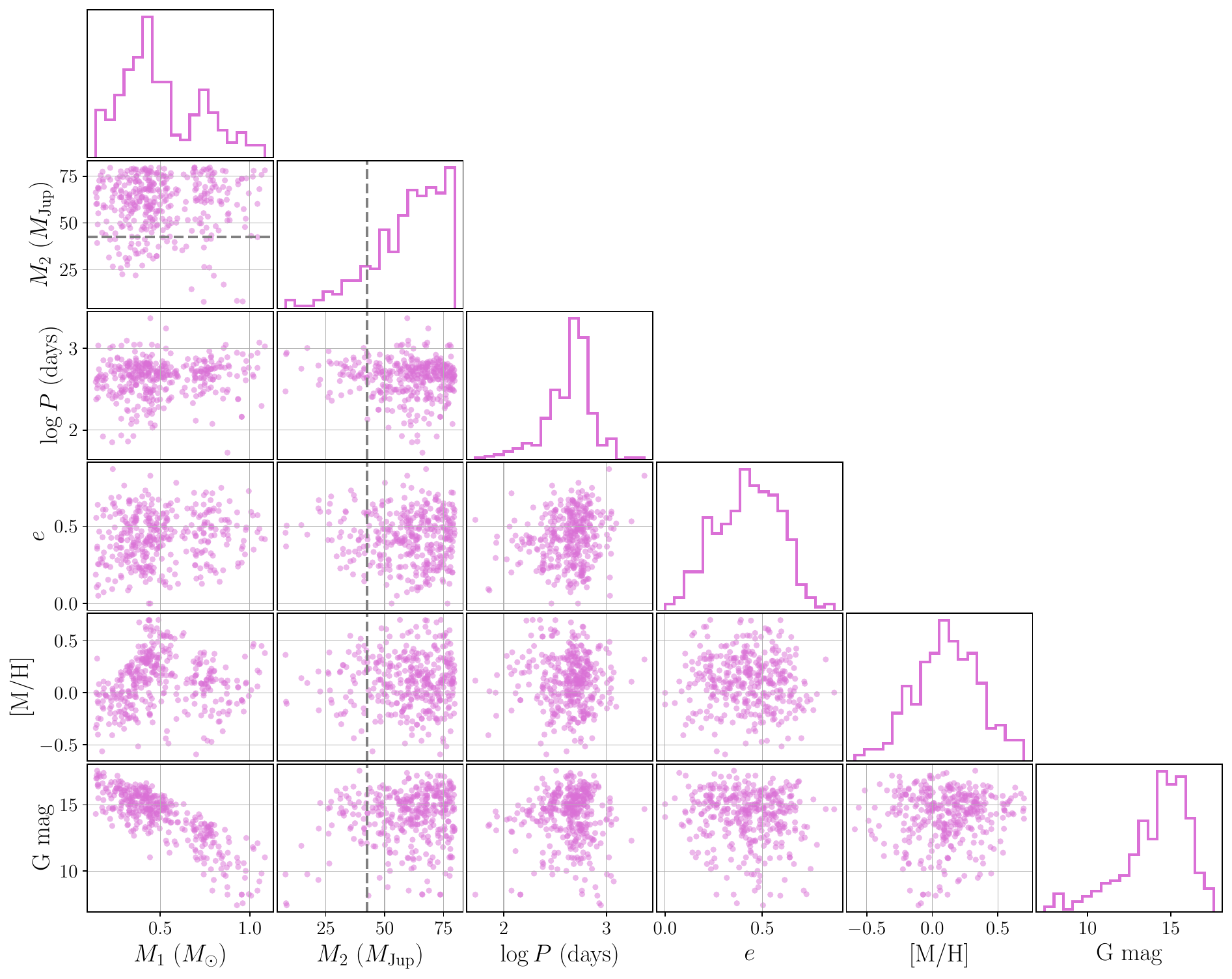}
        \captionof{figure}{Distribution and covariances of relevant parameters. Orbital period $P$ and eccentricity $e$ are obtained directly from the \gaia astrometric solution. The primary mass $M_1$ and metallicity $\mathrm{[M/H]}$ are computed by \cite{Li+2025} using their XP-spectra emulator. The companion mass $M_2$ is computed from the \gaia astrometric solution using the computed primary mass and assuming the companion negligibly contributes to the flux. \gaia apparent G band magnitudes are reported directly from \tgaia. Dashed grey line at $42.5\mjup$ corresponds to the claimed population division by \cite{MaGe14}.}
        \label{fig:SampleCorner}
    \end{minipage}
    \hfill
    \begin{minipage}[t]{0.48\textwidth}
    \vspace{0pt}
        \centering
        \includegraphics[width=\linewidth]{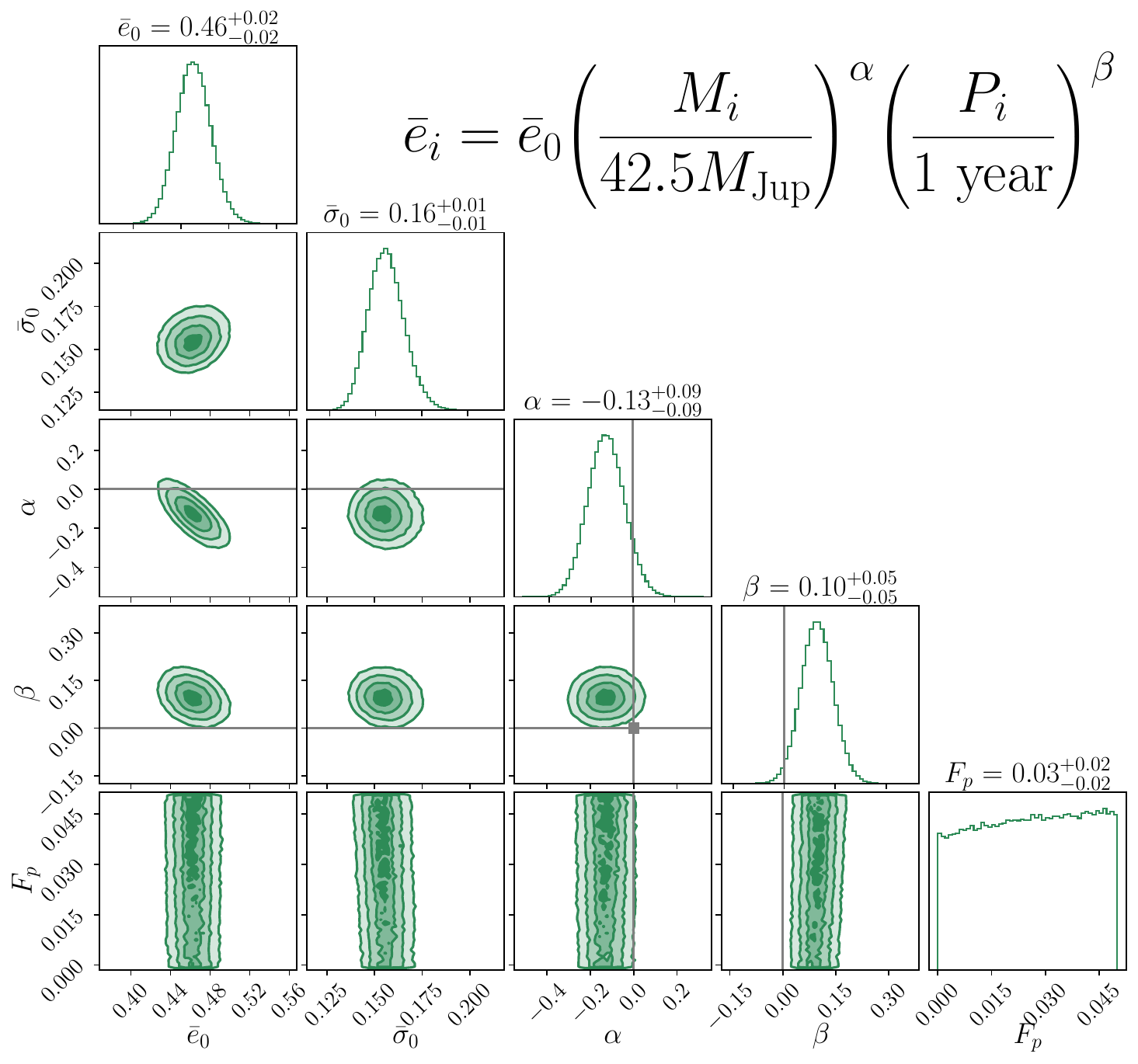}
        \captionof{figure}{BD mass and period dependence on the eccentricity. We create a parametrised model to constrain the peak $\bar{e}$ and width $\sigma_0$ of the eccentricity distribution, where the peak depends on the mass and orbital period of the brown dwarf, with the functional form shown in the figure. The resulting posteriors and covariances of the Monte Carlo Markov Chains are shown in the central panels. The best-fit parameters are shown in red lines in the top right panels with our sample shown in black scatter points to show that the fits are reasonable.}
        \label{fig:EccentricityFit}
    \end{minipage}

    \vspace{0.5cm}

    \includegraphics[width=0.85\textwidth]{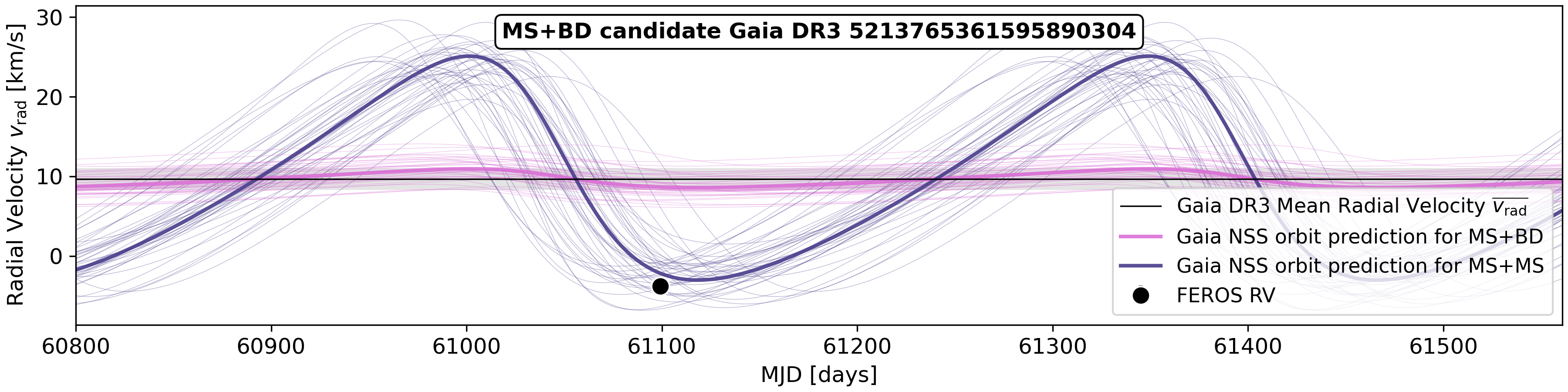}

    \includegraphics[width=0.85\textwidth]{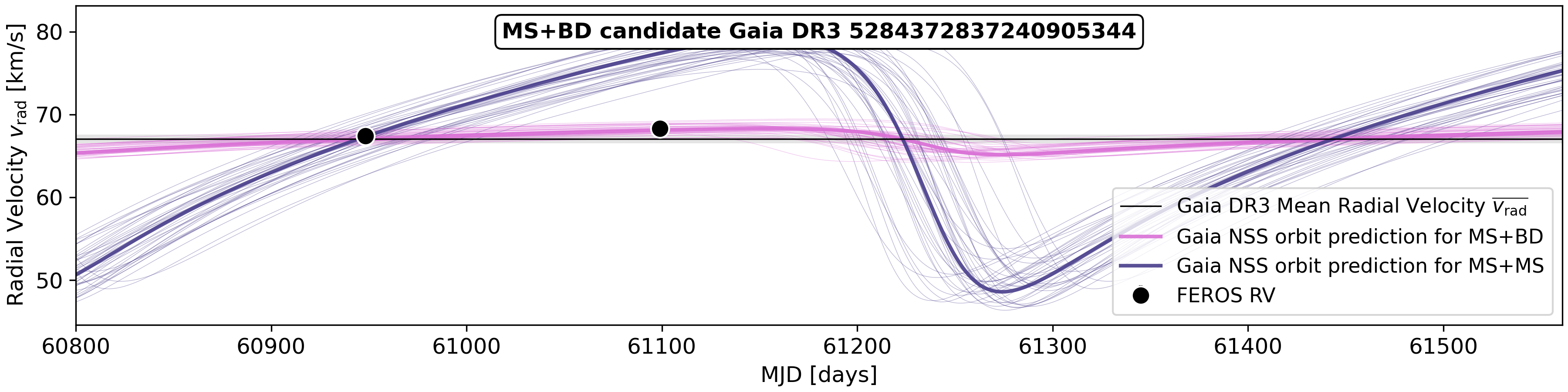}

    \captionof{figure}{Examples of follow up being consistent with an SB2 (top) and BD companion (bottom). BD and SB2 orbits are bootstrapped as outlined in Sect.~\ref{sec:Sample} to sample possible orbits from the covariance matrix under the assumption the companion is low mass (pink) or a near-twin (purple) where a mass-luminosity relationship $M\propto L^4$ is assumed, and the corresponding BD and twin companions are determined from the sampled photocentre semi-major axis.}
    \label{fig:follow-up}

\end{figure*}

We report the likelihood difference as a $\Delta\chi^2$ between values of $(P_\text{max},\delta)$ and the null hypothesis $\delta=0$ for a grid of combinations of the parameters, and show the confidence intervals in Fig.~\ref{fig:DesertLikelihood}.
The result is that the null hypothesis is always preferred, but can only be ruled out at high confidence beyond ${\sim}150$\,days.
This provides no convincing evidence for this flavour of brown dwarf desert in our sample, and that the brown dwarf mass distribution is best explained by being uniform (Fig.~\ref{fig:FiducialComparison}) for the M-, K- and G- type primaries in the sample (Fig.~\ref{fig:SampleCorner}).

\section{The MS-BD sample properties} \label{app:followups}

\begin{table*}
    \caption{All confirmed brown dwarfs, planets, and false positives from other surveys in our sample with orbital parameters derived from \tgaia's orbit solution and primary masses and [M/H] assigned by \cite{Li+2025}.}
    \centering
    \begin{tabular}{rccccccl} 
         \hline
         \hline
         \gaia DR3 source ID & $M_1$ $(M_\odot)$ & $M_2$ $(\mjup)$ & $P$ (days) & $e$ & [M/H] & Designation & Citation(s) \\
         \hline

5148853253106611200 & 0.72 & 52.4 & 340 & 0.17 & 0.28 & BD & \cite{Marcussen2026} \\
3097799420565217408 & 0.83 & 46.0 & 443 & 0.60 & -0.28 & BD & \cite{Marcussen2026} \\
2074815898041643520 & 0.36 & 22.0 & 358 & 0.67 & -0.17 & BD & \cite{Stefansson2025} \\
1712614124767394816 & 0.74 & 7.9 & 298 & 0.46 & 0.36 & BD & \cite{Stefansson2025} \\
1457486023639239296 & 0.67 & 14.6 & 564 & 0.51 & 0.14 & BD & \cite{Stefansson2025} \\
1761619456801656064 & 0.73 & 45.3 & 281 & 0.33 & 0.04 & BD & \cite{Barbato2026} \\
1543883104727682176 & 0.83 & 61.8 & 593 & 0.14 & -0.27 & BD & \cite{Barbato2026} \\
1518957932040718464 & 0.69 & 52.0 & 501 & 0.25 & 0.16 & BD & \cite{Barbato2026} \\
3087990814774129920 & 0.84 & 61.4 & 723 & 0.75 & -0.07 & BD & \cite{Barbato2026} \\
1853704144046204032 & 0.75 & 26.5 & 688 & 0.51 & 0.39 & BD & \cite{Barbato2026} \\
1878822452815621120 & 0.85 & 17.1 & 1009 & 0.45 & 0.09 & BD & \cite{Sahlmann2025} \\
2540855308890440064 & 0.93 & 70.7 & 240 & 0.26 & 0.05 & BD & \cite{Sahlmann2025} \\
685029558383335168\tablefootmark{\textsuperscript{\textdagger}} & 0.79 & 68.8 & 119 & 0.71 & -0.17 & BD & \cite{Wilson2016} \\
855523714036230016\tablefootmark{\textsuperscript{\textdagger}} & 0.95 & 73.7 & 145 & 0.26 & -0.07 & BD & \cite{Wilson2016} \\
3750881083756656128\tablefootmark{\textsuperscript{\textdagger}} & 0.97 & 47.9 & 500 & 0.32 & 0.35 & BD & \cite{Dalal2021} \\
3309006602007842048\tablefootmark{\textsuperscript{\textdagger}} & 1.01 & 43.1 & 815 & 0.59 & 0.03 & BD & \cite{Dalal2021} \\
2778298280881817984\tablefootmark{\textsuperscript{\textdagger}} & 1.04 & 60.2 & 577 & 0.46 & 0.06 & BD & \cite{Dalal2021} \\
2651390587219807744\tablefootmark{\textsuperscript{\textdagger}} & 0.87 & 51.3 & 780 & 0.48 & 0.06 & BD & \cite{Dalal2021} \\
5563001178343925376\tablefootmark{\textsuperscript{\textdagger}} & 0.88 & 66.1 & 53 & 0.54 & 0.12 & BD & \cite{Kiefer2019} \\
824461960796102528\tablefootmark{\textsuperscript{\textdagger}} & 0.99 & 65.9 & 579 & 0.73 & -0.03 & BD & \cite{Sahlmann2011} \\
5999024986946599808 & 0.85 & 69.3 & 243 & 0.22 & -0.18 & BD & \cite{Unger2023} \\
5776191054129718912 & 0.93 & 72.5 & 238 & 0.24 & 0.06 & BD & \cite{Stevenson+23} \\
6071451157905853312 & 0.71 & 62.2 & 530 & 0.34 & 0.33 & BD & \cite{Stevenson+23} \\
1644373176107337344 & 0.84 & 60.1 & 591 & 0.54 & 0.12 & BD & \cite{Stevenson+23} \\
2111646543961513728 & 0.15 & 39.3 & 324 & 0.35 & 0.33 & BD & \cite{Stevenson+23} \\
1390358021264524544 & 0.73 & 64.6 & 565 & 0.49 & -0.22 & BD & \cite{Stevenson+23} \\
4854976609869179648 & 0.98 & 52.3 & 407 & 0.48 & 0.09 & BD & \cite{Stevenson+23} \\
5501729312336121856 & 0.70 & 63.3 & 130 & 0.44 & -0.00 & BD & \cite{Stevenson+23} \\
4062446910648807168 & 0.80 & 21.8 & 615 & 0.62 & 0.20 & BD & \cite{Marcussen2023} \\
637329067477530368 & 0.96 & 8.1 & 851 & 0.37 & -0.09 & BD & \cite{Marcussen2023} \\
5855730584310531200 & 0.93 & 8.4 & 882 & 0.50 & -0.11 & BD & \cite{Marcussen2023} \\
5085864568417061120 & 0.47 & 17.0 & 260 & 0.42 & 0.41 & SB2 & \cite{Stefansson2025} \\
73648110622521600 & 0.25 & 23.3 & 491 & 0.23 & 0.05 & SB2 & \cite{Stefansson2025} \\
2824801747222539648 & 0.46 & 27.7 & 560 & 0.13 & 0.14 & SB2 & \cite{Stefansson2025} \\
3676303512147120512 & 0.24 & 19.7 & 575 & 0.50 & 0.13 & SB2 & \cite{Stefansson2025} \\
5612039087715504640 & 0.29 & 15.0 & 592 & 0.21 & 0.10 & SB2 & \cite{Stefansson2025} \\
4553792062596291072 & 1.08 & 51.0 & 989 & 0.64 & 0.29 & SB2 & \cite{Barbato2026} \\
5213765361595890304 & 0.44 & 24.4 & 348 & 0.28 & 0.52 & SB2 & This work \\
2884087104955208064 & 0.82 & 13.1 & 827 & 0.07 & 0.06 & Single & \cite{Stefansson2025} \\
246890014559489792 & 0.37 & 7.1 & 825 & 0.52 & 0.33 & Single & \cite{Stefansson2025} \\
405316961377489792 & 0.28 & 24.5 & 677 & 0.69 & 0.02 & Single & \cite{Stefansson2025} \\
2446599193562312320 & 0.51 & 28.5 & 854 & 0.43 & 0.12 & Single & \cite{Stefansson2025} \\
         \hline
    \end{tabular}
    \tablefoot{\tablefoottext{\textdagger}{Found before \gaia DR3 \citep{GaiaCollaboration2023}}}
    \label{tab:confirmedbds}
\end{table*}

\begin{table*}
    \caption{All sources followed-up by FEROS in this work}
    \centering
    \begin{tabular}{rccccccl} 
         \hline
         \hline
         \gaia DR3 source ID & $M_1$ $(M_\odot)$ & $M_2$ $(\mjup)$ & $P$ (days) & $e$ & [M/H] & Notes \\
         \hline
1745837139910836096 & 0.76 & 41.2 & 550 & 0.59 & 0.29 & - \\
1761619456801656064 & 0.73 & 45.3 & 281 & 0.33 & 0.04 & Confirmed by \cite{Barbato2026} \\
2457361312158947328 & 0.70 & 68.4 & 655 & 0.16 & 0.17 & - \\
2964776930424653184 & 0.65 & 51.6 & 740 & 0.46 & -0.21 & - \\
3087990814774129920 & 0.84 & 61.4 & 723 & 0.75 & -0.07 & Confirmed by \cite{Barbato2026} \\
3097799420565217408 & 0.83 & 46.0 & 443 & 0.60 & -0.28 & Confirmed by \cite{Marcussen2026} \\
3513138563565911168 & 0.80 & 70.7 & 566 & 0.21 & 0.21 & - \\
4541326418238161792 & 0.75 & 63.3 & 355 & 0.41 & 0.09 & - \\
4713094873506349056 & 0.78 & 67.1 & 553 & 0.67 & 0.13 & Consistent with BD companion \\
4767186928583729792 & 0.62 & 35.4 & 293 & 0.44 & 0.55 & - \\
4811480063016969344 & 0.74 & 36.9 & 726 & 0.70 & -0.44 & - \\
4811689386840344448 & 0.78 & 72.7 & 419 & 0.21 & 0.02 & - \\
4875654059901550848 & 0.77 & 53.4 & 521 & 0.40 & 0.06 & - \\
4912578374781435520 & 0.83 & 67.6 & 656 & 0.64 & -0.20 & - \\
5130193441271871616 & 0.80 & 76.9 & 451 & 0.44 & 0.37 & - \\
5196317344716068224 & 0.59 & 50.1 & 534 & 0.78 & -0.01 & - \\
5213765361595890304 & 0.44 & 24.4 & 348 & 0.28 & 0.52 & SB2 \\
5220375041387610880 & 0.77 & 26.2 & 922 & 0.59 & 0.18 & - \\
5251990536204769920 & 0.76 & 47.5 & 599 & 0.24 & 0.13 & Consistent with BD companion \\
5284372837240905344 & 0.58 & 39.8 & 499 & 0.46 & 0.52 & Consistent with BD companion \\
5387119346996430592 & 0.93 & 59.7 & 459 & 0.59 & 0.12 & - \\
5510211735308380544 & 0.77 & 75.4 & 319 & 0.60 & 0.14 & - \\
5591328892847955968 & 0.81 & 48.3 & 576 & 0.20 & 0.13 & - \\
5612227138561549696 & 0.88 & 54.5 & 312 & 0.44 & -0.21 & - \\
5776759639080398720 & 0.74 & 55.2 & 495 & 0.45 & -0.00 & Consistent with BD companion \\
5845391184955598976 & 0.86 & 72.8 & 578 & 0.54 & 0.09 & - \\
5927476672953974784 & 0.42 & 35.7 & 456 & 0.73 & -0.08 & - \\
6383984898525990272 & 0.55 & 48.8 & 512 & 0.63 & 0.41 & - \\
6506582811548406528 & 0.75 & 69.3 & 694 & 0.34 & 0.08 & Consistent with BD companion \\
6522700384998507264 & 0.88 & 55.2 & 618 & 0.42 & -0.10 & - \\
         \hline
    \end{tabular}
    \tablefoot{Most sources cannot be confirmed due to low radial velocity amplitudes of a BD companion. Those labelled as consistent show radial velocities broadly inconsistent both with zero and a stellar companion.}
    \label{tab:followedupbds}
\end{table*}

\end{appendix}

\end{document}